\documentclass[runningheads]{llncs}
\usepackage[T1]{fontenc}
\usepackage{url}
\usepackage{subcaption}
\usepackage{graphicx}
\usepackage{comment}
\usepackage{booktabs}
\usepackage{upquote} 
\usepackage{listings}
\usepackage{xcolor}
\usepackage{amsmath}
\definecolor{codebg}{RGB}{248,248,250}
\definecolor{codeframe}{RGB}{220,220,220}
\definecolor{codenumbers}{RGB}{130,130,130}
\definecolor{placeholder}{RGB}{90,90,160}
\definecolor{@headline}{rgb}{0.48,0.19,0.64}

\newcommand{\wraparrow}{\(\hookrightarrow\)}

\lstdefinelanguage{gherkin}{
  morekeywords={Feature,Background,Scenario,Scenario Outline,Examples,Given,When,Then,And,But},
  sensitive=false,
  morecomment=[l]{\#},
}

\lstdefinelanguage{json}{
  alsoletter=-,
  morestring=[b]",
  morecomment=[l]{//},
  morecomment=[s]{/*}{*/},
  literate=
   *{:}{{\textcolor{black}{:}}}{1}
    {,}{{\textcolor{black}{,}}}{1}
    {\{}{{\textcolor{black}{\{}}}{1}
    {\}}{{\textcolor{black}{\}}}}{1}
    {[}{{\textcolor{black}{[}}}{1}
    {]}{{\textcolor{black}{]}}}{1}
}

\lstdefinestyle{base-style}{
  backgroundcolor=\color{codebg},
  frame=single,
  rulecolor=\color{codeframe},
  frameround=tttt,
  numbers=left,
  numberstyle=\color{codenumbers}\footnotesize,
  numbersep=8pt,
  xleftmargin=0em,
  framexleftmargin=0em,
  framesep=2.5pt,
  showstringspaces=false,
  breaklines=true,
  breakatwhitespace=true,
  postbreak=\mbox{\textcolor{codenumbers}{\wraparrow}\space},
  tabsize=2,
  captionpos=b,
  columns=fullflexible,
  keepspaces=true,
  breakindent=0pt,
  linewidth=\linewidth,
  basicstyle=\ttfamily\footnotesize,
  upquote=true
}

\lstdefinestyle{gherkin-style}{
  style=base-style,
  language=gherkin,
  numbers=none,
  xleftmargin=0.0em,
  framexleftmargin=0.6em,
  basicstyle=\ttfamily\scriptsize,
  keywordstyle=\bfseries,
  commentstyle=\itshape\color{codenumbers},
  breakatwhitespace=true,
  breakindent=2.2em,
  postbreak=\mbox{\textcolor{codenumbers}{\(\hookrightarrow\)}\hspace{0.4em}},
  moredelim=**[is][\color{codenumbers}]{[}{]},
  alsoletter=._@,
  emph={@Req_CPDS_00_1,@Req_CPDS_00_2,@Req_CPDS_01_1,@Req_CPDS_04_1,
        @Req_CPDS_04_2,@Req_CPDS_04_3,@Req_CPDS_04_4},
  emphstyle=\color{@headline}\bfseries,
}

\lstdefinestyle{json-style}{
  style=base-style,
  language=json
}

\lstdefinestyle{prompt-style}{
  style=base-style,
  language={},
  numbers=none,
  basicstyle=\ttfamily\scriptsize, 
  breaklines=true,
  breakindent=0pt,
  postbreak=\mbox{\hspace{0.4em}},
  moredelim=**[s][\color{codenumbers}]{[}{]},   
  moredelim=**[s][\color{placeholder}]{\{}{\}}  
}

\lstdefinestyle{python-style}{
  style=base-style,
  language=Python,
  numbers=none,          
  xleftmargin=0.6em,           
  framexleftmargin=0.6em,
  basicstyle=\ttfamily\footnotesize,
  keywordstyle=\bfseries,
  commentstyle=\itshape\color{codenumbers},
  stringstyle=\color{black},
  breakatwhitespace=true,
  breakindent=2em,
  upquote=true,
  moredelim=**[s][\color{codenumbers}]{[}{]},
}

\lstdefinestyle{console-style}{
  style=base-style,
  language={},             
  numbers=none,
  basicstyle=\ttfamily\scriptsize,
  breaklines=true,
  breakatwhitespace=true,
  keepspaces=true,
  postbreak=\mbox{\hspace{0.4em}},  
  emph={APPL:,MAIN:,Feature:,Scenario:,Behave,Exit,STDOUT,STDERR},
  emphstyle=\color{codenumbers}
}

\lstdefinestyle{requirements}{
  style=base-style,
  language={},
  numbers=none,
  xleftmargin=0.6em,
  framexleftmargin=0.6em,
  basicstyle=\ttfamily\scriptsize, 
  breaklines=true,
  breakindent=0pt,
  postbreak=\mbox{\hspace{0.4em}},
  alsoletter=._,
  emph={Req_CPDS_04,Req_CPDS_04.1,Req_CPDS_04.2,Req_CPDS_04.3,Req_CPDS_04.4},
  emphstyle=\color{codenumbers}\bfseries
}

\begin{document}
\title{Req2Road: A GenAI Pipeline for SDV Test Artifact Generation and On-Vehicle Execution}
\titlerunning{Req2Road}
%
\author{Denesa Zyberaj\inst{1,2}\orcidID{0009-0009-4207-8723} \and
Lukasz Mazur\inst{3}\orcidID{0009-0002-0997-0997} \and
Pascal Hirmer\inst{1}\orcidID{0000-0002-2656-0095} \and 
Nenad Petrovic\inst{3}\orcidID{0000-0003-2264-7369} \and 
Marco Aiello \inst{2}\orcidID{0000-0002-0764-2124} \and 
Alois Knoll\inst{3}\orcidID{0000-0003-4840-076X}}
\authorrunning{Zyberaj et al.}
%
\institute{Mercedes-Benz AG, Bela-Barenyi-Straße, 71059 Sindelfingen, Germany\\
\email{\{denesa.zyberaj, pascal.hirmer\}@mercedes-benz.com}
\and
Institute for Architecture of Application Systems, University of Stuttgart,\\
Universitätsstraße 38, 70569 Stuttgart, Germany\\
\email{marco.aiello@iaas.uni-stuttgart.de} \and
Institute for Robotics, Artificial Intelligence and Embedded Systems, Technical University of Munich, Boltzmannstraße 3, 85748 Garching, Germany\\
\email{\{ lukasz.mazur, nenad.petrovic, k\}@tum.de}}
\maketitle              
\begin{abstract}
Testing functionality in Software-Defined Vehicles is challenging because requirements are written in natural language, specifications combine text, tables, and diagrams, while test assets are scattered across heterogeneous toolchains. Large Language Models and Vision-Language Models are used to extract signals and behavioral logic to automatically generate Gherkin scenarios, which are then converted into runnable test scripts. The Vehicle Signal Specification (VSS) integration standardizes signal references, supporting portability across subsystems and test benches. The pipeline uses retrieval-augmented generation to preselect candidate VSS signals before mapping. We evaluate the approach on the safety-relevant Child Presence Detection System, executing the generated tests in a virtual environment and on an actual vehicle. Our evaluation covers Gherkin validity, VSS mapping quality, and end-to-end executability. Results show that 32 of 36 requirements (89\%) can be transformed into executable scenarios in our setting, while human review and targeted substitutions remain necessary. This paper is a feasibility and architectural demonstration of an end-to-end requirements-to-test pipeline for SDV subsystems, evaluated on a CPDS case in simulation and Vehicle-in-the-Loop settings.
\end{abstract}

\begin{figure}[t]
    \centering
    \includegraphics[width=0.7\linewidth]{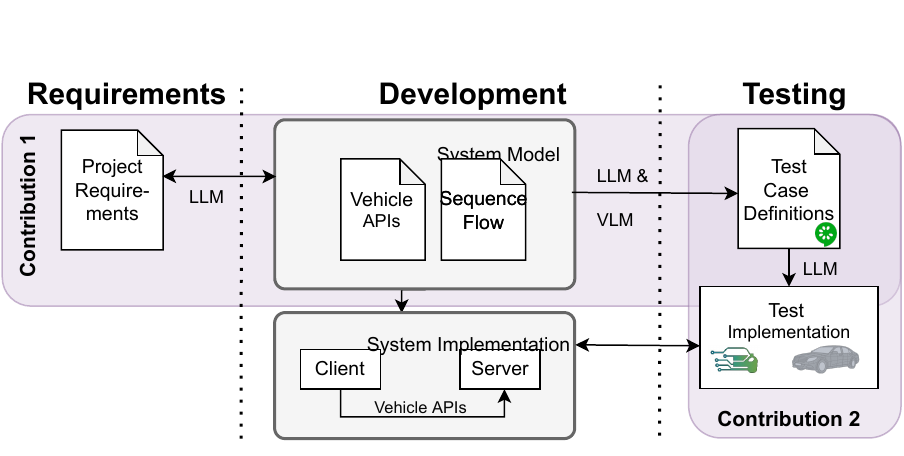}
    \caption{Overview of our proposed GenAI-based test generation.}
    \label{fig:intro}
\end{figure}
\section{Introduction}
The transition to Software-Defined Vehicles (SDVs) increases software complexity, leading to higher test volume, greater variability, and increased strain on systematic verification and validation. Despite established standards such as \mbox{ASPICE}, automotive test planning often remains inconsistent~\cite{Lami2016}, manual~\cite{Juhke2020}, and poorly aligned with evolving requirements~\cite{Ambrosio2017}. Requirements expressed in natural language further complicate automated test derivation~\cite{bruel2020} and limit traceability across artifacts~\cite{Garousi2017}. In practice, interoperability gaps from heterogeneous tools, mixed-media specifications, and inconsistent signal definitions hinder automation, portability, and reuse across test benches and subsystems~\cite{Zyberaj2025}.

Generative AI provides a practical path to transform ambiguous inputs into structured, executable testing artifacts. In this work, we use Large Language Models (LLMs) and Vision-Language Models (VLMs) to extract signals and behavioral logic from requirements text, tables, and diagrams, also illustrated in Figure~\ref{fig:intro}. The models generate Gherkin~\cite{cucumber_gherkin_docs} scenarios, which serve as input for runnable Python test scripts. We integrate the Vehicle Signal Specification (VSS)~\cite{covesa_vss} to avoid hard-coded signals, promote hardware abstraction, and enable test portability across subsystems and benches. We target the digital.auto playground~\cite{digitalauto} as a rapid execution environment so that teams can validate requirement semantics and system integration, and we further demonstrate on-vehicle execution of the generated CPDS test scripts.
Req2Road is designed to translate requirements into runnable scenarios, standardize signal references via VSS, and enable artifact reuse through standardized representations; this paper evaluates these goals as a feasibility and architectural demonstration on the safety-relevant CPDS case, including execution in the digital.auto playground and on an actual vehicle.
We present Req2Road as an architectural method and feasibility study. The observed efficiency and reuse benefits in this case study are reported as indicative and serve to motivate future comparative studies.
All artifacts for our scenario, i.e., prompts, Gherkin feature files, VSS mappings, and the generated Python test scripts, are provided as supplementary material. They are publicly available as a digital.auto vehicle model titled \textit{Req2Road: A GenAI Pipeline for SDV Test Artifact Generation and On-Vehicle Execution}.\footnote{Req2Road project artifacts:\\ \url{https://playground.digital.auto/model/69246d3cd327158aa9737ee3}}


\section{Background and Related Work}    
\label{background_relatedwork}
Requirements engineering (RE) studies repeatedly report that many problems in software projects originate from mistakes during the requirements phase. Tukur et al.\ synthesize evidence from 70 primary studies and a practitioner survey and identify a set of recurring RE challenges, including unclear or evolving system requirements, limited domain knowledge on the development side, communication gaps, and difficulties when formalizing requirements that are initially expressed only in natural language~\cite{Tukur2021}. In the automotive context, Juhnke et al.\ observe similar problems at the test specification level and argue that preparing and maintaining high-quality test cases is labor-intensive, especially when many stakeholders and rapidly changing requirements are involved~\cite{Juhke2020}.

Traditional solutions rely on model- and scenario-based test generation, where tests are derived from structured artifacts such as UML or SysML models, Stateflow diagrams, or domain-specific languages rather than directly from natural-language requirements, tables, and diagrams. This shifts much of the effort to building and maintaining these models, while heterogeneous requirements artifacts remain underused as a direct source for test cases.

Several approaches instantiate this strategy for automotive software~\cite{Shin2018,Tekaya2014}. Shin and Lim generate unit, integration, and Hardware-in-the-Loop (HiL) test cases fully automatically from UML state and class diagrams for a power-window control module~\cite{Shin2018}, relying on static model analysis, custom parsers for UML action code, and search-based recombination of unit tests under the assumption that detailed behavioral models are already available. Tekaya et al.\ propose MB-ATG, where structural coverage criteria from ISO~26262 are encoded as properties on Simulink/Stateflow models and processed by Simulink Design Verifier to obtain minimal test suites for engine control functions~\cite{Tekaya2014}. Both approaches focus on deriving low-level test vectors from design models for ECU-oriented Model-in-the-Loop/ Software-in-the-Loop/ Hardware-in-the-Loop (MiL/SiL/HiL) environments.

Beyond classical model-based techniques, several recent approaches use AI assistance to generate automotive test artifacts~\cite{WynnWilliams2025,Karlsson2025,Lebioda2025,Zhang2024,Tan2023}. Wynn-Williams et al.~\cite{WynnWilliams2025} investigate generative AI to translate informal test case specifications from IBM Rational Quality Manager into executable \textit{ecu.test} scripts for an automotive HiL environment, while Karlsson et al.~\cite{Karlsson2025} study the use of GitHub Copilot to generate C\# HiL test cases within Volvo Trucks' Test Automation Framework. Both approaches start from existing test descriptions and requirements and focus on producing low-level, framework-specific scripts that can be executed on ECU-oriented HiL benches. Lebioda et al.~\cite{Lebioda2025} present a pipeline that converts high-level requirements for Autonomous Emergency Braking (AEB) into CARLA configuration code and introduce a three-way classification of requirements (direct, indirect, abstract) to capture how easily they can be mapped to concrete configuration parameters. ChatScene uses LLM-based agents to generate semi-formal driving scenarios and map them to Scenic code snippets via a snippet database~\cite{Zhang2024}, while LCTGen converts natural-language test descriptions into a compressed internal representation that is later expanded into concrete tests~\cite{Tan2023}. Collectively, these approaches focus on translating textual descriptions into simulator- or framework-specific scenario configurations and code snippets for particular back-end tools. In contrast, Req2Road operates earlier in the lifecycle and at a higher level of abstraction by starting from heterogeneous system specification artifacts (requirements text, tables, UML diagrams) and using GenAI to derive scenario-based test artifacts, which are evaluated in SiL and Vehicle-in-the-Loop (ViL) setups rather than framework-specific HiL scripts or simulator configurations.

To support such GenAI-based pipelines in an SDV setting, we build on three existing foundations. Behavior-driven development (BDD) notations, such as Gherkin, capture scenario-based tests in a form that is readable for non-technical stakeholders while remaining directly executable in test frameworks like Cucumber, Behave, or SpecFlow~\cite{dosSantos2018}. This makes Gherkin a natural target for automation from natural-language requirements. The Vehicle Signal Specification defines a vendor-neutral, hierarchically structured catalog of vehicle signals and standard get/set/subscribe APIs that have been adopted in several SDV toolchains to decouple application logic and tests from proprietary ECU interfaces~\cite{aust2022vehicle}. Finally, the digital.auto SDV playground combines VSS-based signal access with a cloud-based, containerized execution environment to support rapid SiL experimentation and providing a migration path toward ViL setups~\cite{aust2022vehicle,slama2023digitalauto}. 

\section{Req2Road}
\label{approach}
\emph{Req2Road} (Requirements-to-Road) is a GenAI-based process that converts specification artifacts into executable tests for SDVs. We first describe the pipeline that uses these models, detailing its stages and the prompt designs employed at each stage, and then evaluate representative LLMs and VLMs on automotive testing tasks. 
In this paper, ‘executable’ is defined at three levels: (1) executable specification: the generated Gherkin feature parses under the BDD framework (Behave) and executes without syntax/runtime errors; (2) executable test implementation: corresponding step definitions and harness code compile/import and execute; (3) executable end-to-end test: the test suite runs against a target backend (SiL or ViL) and produces pass/fail based on explicit oracle checks.

\subsection{GenAI-based Test Pipeline}
\label{req2road_pipeline}
\begin{figure}[t]
    \centering
   \includegraphics[width=1.0\linewidth]{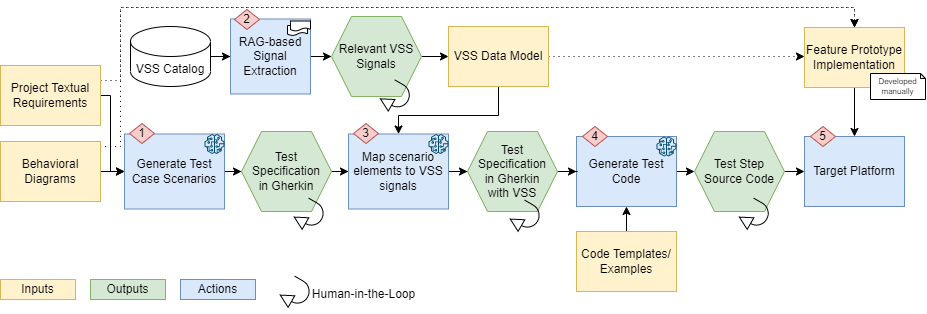}
\caption{Req2Road workflow: requirements to executable test artifacts.}
\label{fig:wf_req2road}
\end{figure}

Figure~\ref{fig:wf_req2road} provides an overview of Req2Road. The pipeline comprises four phases: (i) initial Gherkin scenario generation, (ii) VSS signal mapping, (iii) refined Gherkin scenario generation, and (iv) code generation. It begins with project textual requirements and behavioral diagrams, which, together with the VSS catalog, serve as inputs. The RAG-based signal extraction module identifies relevant VSS signals as an intermediate result and constructs a corresponding VSS data model to guide subsequent stages.

Our process includes two AI-assisted phases of Gherkin scenario generation. In the preliminary phase, initial Gherkin test scenarios are produced from textual and behavioral specifications, providing a basis for Human-in-the-Loop refinement. In the refined phase, these corrected scenarios are mapped to VSS signals through a combination of LLMs and VLMs, enabling both semantic and visual grounding of scenario elements. The result is a VSS-enriched Gherkin test specification, which ensures alignment between test logic and vehicle signal semantics.
Finally, test code is generated using predefined code templates, producing executable test code deployed to the target platform. The feature prototype implementation, developed manually, serves as the system under test. Color coding in Figure~\ref{fig:wf_req2road} differentiates inputs (yellow), outputs (green), and actions (blue), emphasizing the human–AI collaborative pipeline that bridges requirements understanding, model-based signal mapping, and automated test realization.

Before code generation, we aim to reduce hallucinations by extracting only scenario-relevant VSS signals. The VSS mapping phase begins with the hierarchical VSS JSON catalog, where each entry contains fields such as name, data type, and description. A preprocessing script recursively traverses the hierarchy and produces a flattened list of signals in the form \texttt{signalName, sensor/actuator, datatype} (one per line). This list serves as the authoritative reference for valid signals used in downstream prompts.

To handle the large catalog and keep the mapping task tractable, we adopt a Retrieval Augmented Generation (RAG) setup following Zolfaghari et al.~\cite{Zolfaghari2024RAG}. The driving scenario is provided as natural-language text or as a flowchart-like diagram. The scenario and each VSS entry are embedded with a SentenceTransformer encoder, and nearest-neighbor search retrieves a small candidate set. A cross-encoder re-ranks these candidates with the scenario text, and we only keep the top $N$ signals (e.g., $N = 16$) as the shortlist. This retrieval-first workflow confines generation to predefined, valid entries, significantly reducing hallucinated or irrelevant signals compared to prompting directly over the full catalog.

The shortlisted signals are then injected into a compact natural-language prompt that combines the scenario description and the candidate list. The prompt instructs the LLM to select only the relevant signals from this list and to return them as a comma-separated list of VSS paths. The model output is normalized (trimming whitespace, standardizing delimiters, and removing duplicates) to form the final set of mapped signals, which can optionally be compared against CPDS ground truth for evaluation (Section~\ref{req2road_eval}). The complete prompt template is provided in the supplementary material. Once the VSS signals are identified, they are used to enrich the Gherkin scenarios with explicit VSS paths and to guide the subsequent code generation stage.

\subsection{LLM and VLM Evaluation}
\label{req2road_eval}
LLMs support summarization, analysis, and text generation, which makes them suitable for tasks in our pipeline such as mapping requirement-level signals to VSS paths and generating code from test case specifications. For visual artifacts, we leverage VLMs to extract relevant information from UML sequence and state diagrams, building on our prior work~\cite{Petrovic2025}. We consider both hosted commercial models, such as OpenAI’s GPT-4o-mini and GPT-5, as well as locally deployable models with fewer parameters, such as Llama-based open-source variants.

Our evaluation uses a curated set of 36 natural-language requirements derived from the CPDS escalation logic specification of this study (based on Euro NCAP guidance and relevant patents) and consolidated into a stable version for benchmarking. Requirements were reviewed to remove duplicates and inconsistencies in wording before running the experiments. Ground truth for the VSS mapping benchmark was constructed by selecting the expected VSS paths from the VSS catalog for each evaluated CPDS scenario (gold set). The gold set was cross-checked by a second reviewer. For the VLM diagram benchmark, ground truth was defined at the state/transition level by annotating the UML state diagram (8 states, 11 transitions) with the relevant signals from the 16-signal candidate pool; annotations were verified by a second reviewer. These ground-truth labels are used only to benchmark model outputs in this feasibility study.

We evaluate LLMs on three steps that align with the pipeline: Gherkin scenario generation, VSS signal mapping, and Python test script generation. The study uses the CPDS case, targeting the digital.auto playground for SiL testing, and assumes Python as the execution language. For the mapping step, we compare GPT-4o-mini and a local Vicuna model in two candidate-pool settings: a 16-signal shortlist of commonly used signals and the full in-vehicle catalog of 982 signals. Decoding settings are fixed across runs (temperature 0, max tokens 1024, and top-k 25, which selects up to 25 relevant VSS signals) without fine-tuning. 

\begin{table}[t]
\caption{VSS signal mapping on CPDS with LLMs. “Correct” = matches among four ground truth signals; “Additional (FP)” = extra proposals.}
\centering
\small
\renewcommand{\arraystretch}{1.1}
\setlength{\tabcolsep}{6pt}
\begin{tabular}{lccc}
\toprule
\textbf{Model} & \textbf{\shortstack{Candidate\\signals}} & \textbf{\shortstack{Correct/\\Expected}} & \textbf{\shortstack{Additional\\ (FP)}} \\
\midrule
GPT-4o-mini               & 16  & 4/4 & 0  \\
TheBloke/vicuna-7B-1.1-HF & 16  & 4/4 & 7  \\
GPT-4o-mini               & 982 & 2/4 & 4  \\
TheBloke/vicuna-7B-1.1-HF & 982 & 4/4 & 19 \\
\bottomrule
\end{tabular}
\label{tab:llm_signal_mapping}
\end{table}

Results show that limiting the candidate pool improves mapping quality. With 16 candidates, GPT-4o-mini matches the gold set (4/4), while the local Vicuna variant also finds all four but adds seven false positives. With the full catalog (982), both degrade: Vicuna preserves recall but adds 19 false positives; GPT-4o-mini finds 2 of 4 with fewer extras. Hence, we prefilter to a small shortlist of probable VSS signals rather than applying RAG over the full catalog at run time. Without retrieval or prefiltering, neither model reliably selects the correct signals. Although the commercial model outperforms the smaller one, locally deployable models remain valuable when requirements and scenarios are sensitive, consistent with~\cite{Petrovic2025GenAI}; accordingly, our pipeline supports local deployment.

We also compare VLMs for diagram-based VSS signal extraction (Table~\ref{tab:vlm_precision_recall}). The benchmark utilizes a UML state diagram with 8 states and 11 transitions, along with a candidate pool of 16 VSS signals. Among API models, Gemini 2.5 Pro achieves the highest precision and recall; among locally deployable models, Qwen 2.5 VL 72B performs best. For specifications that contain sensitive information, for example, proprietary requirements or user stories, local deployment prevents the sharing of artifacts with external providers.
\begin{table}[t]
\caption{VLM performance on diagram-based VSS signal extraction (CPDS).}
\label{tab:vlm_precision_recall}
\centering
\small
\renewcommand{\arraystretch}{1.1}
\setlength{\tabcolsep}{8pt}
\begin{tabular}{lrr}
\toprule
\textbf{Model} & \textbf{Precision} & \textbf{Recall} \\
\midrule
Gemini 2.5 Pro                         & 0.9556 & 0.9556 \\
GPT-4o                                  & 0.7800 & 0.8667 \\
Grok 3                                  & 0.7622 & 0.7778 \\
Qwen 2.5 VL 72B (local)                 & 0.8433 & 0.8222 \\
Llama 4 Maverick 17B 128E Instruct (local) & 0.7867 & 0.8222 \\
InternVL 3 78B (local)                  & 0.6547 & 0.7111 \\
\bottomrule
\end{tabular}
\end{table}

To assess code generation quality, we adopt the pass@1 and pass@3 metrics as defined by Chen et al.~\cite{chen2021evaluating}, which capture the probability that at least one of $k$ generated completions solves a problem completely. The models receive refined Gherkin scenarios that already include VSS mappings and must generate Python test scripts that satisfy one, two, or three CPDS requirements (\textit{Req\_CPDS\_04.1}, \textit{Req\_CPDS\_05.1}, and \textit{Req\_CPDS\_08.1}) using a shortlist of seven candidate VSS signals. For scenarios derived from one or two requirements, GPT-4o-mini, GPT-4.1, and GPT-5 all achieve pass@3 equal to 1. When all three requirements are combined in a single scenario, only GPT-4.1 reaches pass@3 equal to 1; GPT-4o-mini and GPT-5 fail to produce any fully correct solutions. Manual inspection shows that failures involve either syntax errors (non-existent attributes) or violations of semantic conditions such as incorrect termination logic. These findings suggest that correctness remains sensitive to scenario complexity, and that sampling, execution-based checking, and human review continue to be necessary. Based on this comparison, we use GPT-4.1 for the final test script generation in Req2Road and GPT-4o-mini for lighter-weight tasks such as VSS mapping.

\section{Implementation}
\label{implementation}
Req2Road has been implemented as a prototype CPDS pipeline for test case generation, test script generation, and test execution. We utilize GPT-5 for generating Gherkin test cases, Qwen 2.5 VL 72B as the local VLM for diagram-based signal extraction, and GPT-4o-mini to map VSS paths to the test cases. For generating the Python test scripts, we use GPT-4.1.

\subsection{Use Case: Child Presence Detection System}
\label{usecase} 
A Child Presence Detection System (CPDS) detects and responds to unattended children in parked vehicles. We model a simplified CPDS based on Euro NCAP guidelines~\cite{euroncap} and relevant patents~\cite{patent1,patent2}. 
\begin{figure}[t]
    \centering
    \includegraphics[width=0.7\linewidth]{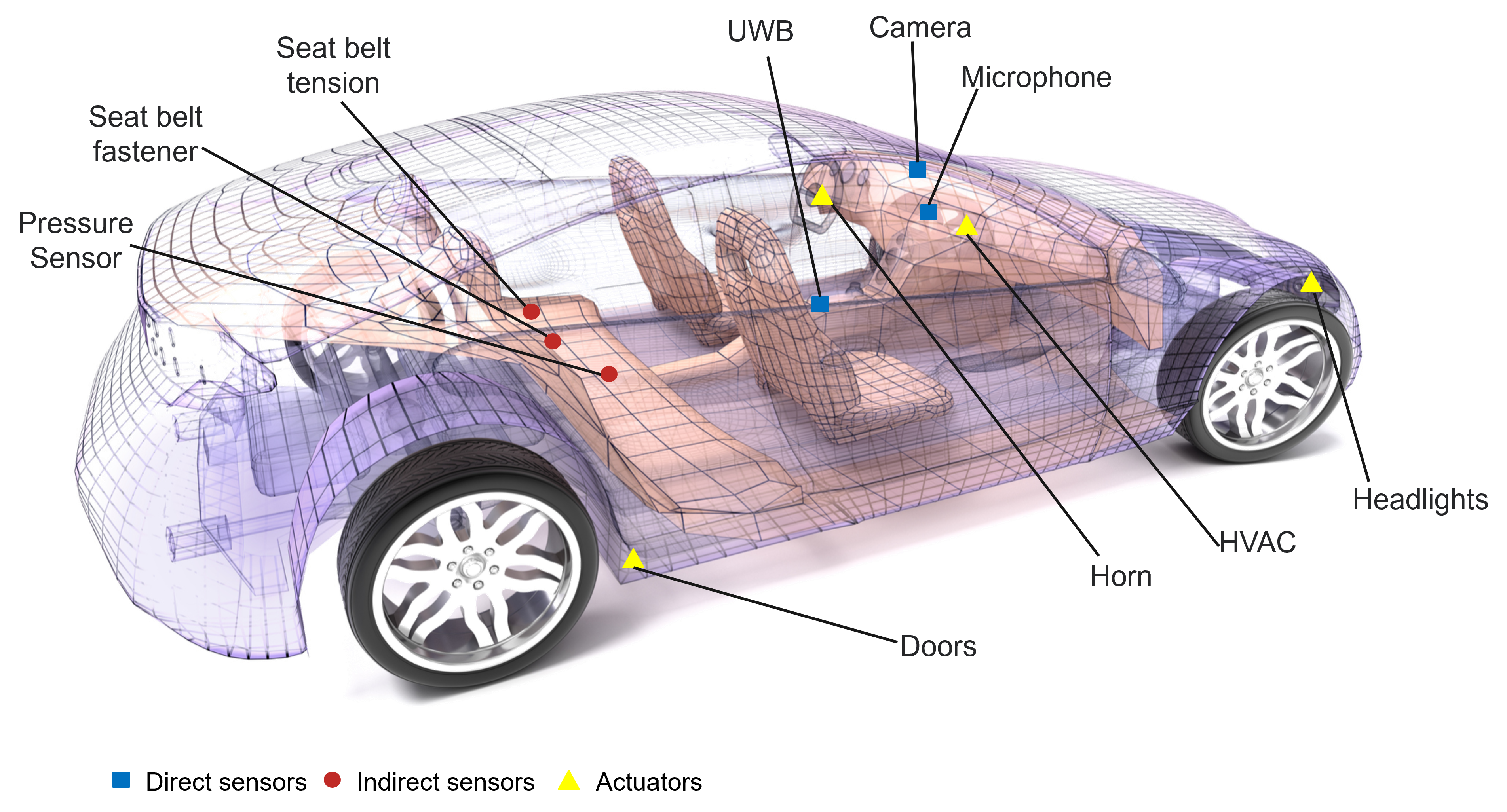}
    \caption{CPDS sensors and actuators considered in this study.}
    \label{fig:vehicle_sensors_actuators}
\end{figure}
The system combines direct in-cabin sensing (e.g., cameras, Ultra-Wideband modules, microphones) with indirect cues (seat-pressure monitors and seatbelt indicators), and actuates the Heating, Ventilation, and Air Conditioning (HVAC) system, driver notifications, exterior lights, horns, and door locks (Figure~\ref{fig:vehicle_sensors_actuators}).

\begin{figure}[t]
    \centering
    \includegraphics[width=1\linewidth]{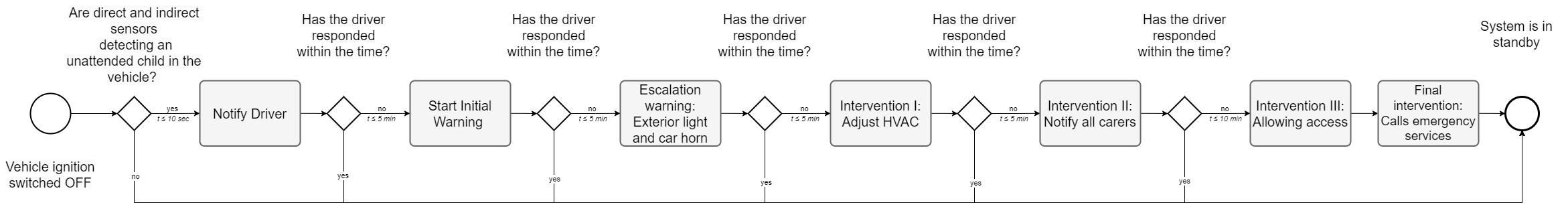}
    \caption{CPDS escalation logic with time-limited escalation stages.}
    \label{fig:cpd}
\end{figure}

After ignition is switched off, CPDS evaluates sensor data within 10 seconds to decide whether a child is unattended (Figure~\ref{fig:cpd}). If a child is detected, the driver is notified and given a 5-minute response window. Missing or invalid responses trigger a time-based escalation through notification and intervention stages, including exterior light and horn activation, HVAC adjustments, caregiver notification, and ultimately contacting emergency services, before the system returns to standby~\cite{Zyberaj2025}.
The escalation process is captured in 36 natural-language requirements. We use Intervention~I (HVAC adjustment) as a running example. The complete requirement set is provided as supplementary material.

\subsection{Test Case Generation}
The first step is to generate test cases in Gherkin syntax from natural language requirements and the escalation flowchart. Listing~\ref{lst:gherkin_input} shows the \textit{HVAC adjustment intervention (Req\_CPDS\_04)}, which we use as a running example for test case generation. We show only representative excerpts due to space constraints; the full artifacts are provided in the supplementary material.

\begin{lstlisting}[style=requirements,
  caption={HVAC adjustment requirements in natural language (excerpt).},
  label={lst:gherkin_input}]
Req_CPDS_04: HVAC Adjustment Intervention
Req_CPDS_04.1: If no valid acknowledgment by T_ACK_2, CPDS shall enter S5 and control HVAC to maintain SAFE_TEMP_RANGE while preserving battery safety limits.
Req_CPDS_04.2: (Battery guard): If traction or 12 V battery state of charge is below a calibrated threshold, CPDS shall prioritize minimal-energy cooling or ventilation and continue escalation. If SOC < SOC_CRIT, Req_CPDS_04.4 takes precedence.
\end{lstlisting}

We used GPT-5 in combination with a VLM to generate Gherkin scenarios from natural language requirements and the flowchart  shown in Fig.~\ref{fig:cpd}.
For the vision-language component, Gemini 2.5 Pro provided the most promising results; however, among the locally deployable models, Qwen 2.5 VL 72B performed best and is thus used in our Req2Road pipeline.
Using locally deployable GenAI models is especially valuable for handling sensitive automotive-specific artifacts, such as requirements and user stories, that should not be shared externally.
The main prompting techniques we used for test case generation are Chain of Thought and Few-Shot Learning. Listing~\ref{lst:prompt_testcasegen} shows the prompt template.

\begin{lstlisting}[style=prompt-style,
  caption={Prompt used for test case generation.}, label={lst:prompt_testcasegen}]
System message:
You are a software testing assistant specialized in automotive safety systems. Your task is to generate structured test cases in Cucumber Gherkin syntax using the Given/When/Then format for a Child Presence Detection System (CPDS). The CPDS identifies unattended children in a parked vehicle and triggers time-based safety interventions.
Each test case should:
- Correspond to a specific step or transition in the escalation flowchart.
- Reflect the timing conditions explicitly (for example, "within 5 minutes").
- Align with the functional requirements.
If any input appears ambiguous or inconsistent, ask for clarification beforehand.
Human message:
--- Natural language requirements describing system behavior ---
{requirements_list}
--- Flowchart illustrating the time-based escalation logic ---
{flowchart}
--- Gherkin syntax reference ---
{gherkin_example}
\end{lstlisting}

We manually reviewed all generated test cases in a human-in-the-loop process to check for semantic and syntactic errors. Req2Road automates the initial draft of (i) Gherkin scenarios, (ii) VSS path selections, and (iii) Python step skeletons, while human reviewers validate intent, thresholds, and signal semantics. In our CPDS case study, based on recorded review notes, review and correction required approximately and correction required approximately 1--5 min per requirement for Gherkin, 1--3 min per scenario for VSS mapping (shortlist-based), and 5--20 min per scenario for code adaptation and oracle checks, depending on scenario complexity and backend configuration. This review assumes (a) basic CPDS domain familiarity, (b) ability to interpret VSS signal descriptions and naming conventions, and (c) practical test automation skills in Python/Behave and KUKSA-based signal access.

Overall, 32 of 36 requirements (89\%) were transformed into executable Gherkin scenarios without modification; the remaining four required targeted adjustments. Most issues occurred in poorly specified edge cases and in requirements with OR conditions that did not explicitly enumerate all options. To avoid overwhelming the model and to simplify checking, we separated initial Gherkin test case generation from a refinement step that injects the corresponding VSS signal paths into the scenarios. For the VSS mapping step, we used GPT-4o-mini as the LLM backend, as it provided the best trade-off between correct matches and false positives on the curated candidate pool (Section~\ref{req2road_eval}).
Listing~\ref{lst:gherkin_scenario} shows an excerpt of the generated Gherkin scenarios with VSS paths for the HVAC adjustment intervention (Req\_CPDS\_04) as input to the test script generator.

\begin{figure*}[t]
\centering
\begin{lstlisting}[style=gherkin-style,
  caption={Generated Gherkin scenarios for Intervention I (excerpt).}, label={lst:gherkin_scenario}]
Feature: CPDS HVAC adjustment intervention (Intervention I, Req_CPDS_04)
  # High-level context: child confirmed, escalation already running.
 @Req_CPDS_04_1 @Req_CPDS_04_2
  Scenario Outline: CPDS_04_1_activate_HVAC_and_apply_battery_guard
    Given traction battery SOC is <SOC_TRACTION_PCT> percent [Req_CPDS_04.2]
    And 12 V battery SOC is <SOC_12V_PCT> percent [Req_CPDS_04.2]
    And a battery guard threshold is configured as <BAT_GUARD_PCT> percent [Req_CPDS_04.2]
    And no valid driver acknowledgment has been received within T_ACK_2 [Req_CPDS_02.2]

    When T_ACK_2 elapses with no valid acknowledgment [Req_CPDS_03.1]
    Then CPDS state becomes S5 HVAC Intervention [Req_CPDS_04.1]
    And Vehicle.Cabin.Infotainment.HVAC.AutoOverrideActive is set to true [Req_CPDS_04.1]
    And Vehicle.Cabin.HVAC.CabinTemperature is controlled into SAFE_TEMP_RANGE <SAFE_MIN_C> to <SAFE_MAX_C> [Req_CPDS_04.1]
    And HVAC operation respects battery safety limits [Req_CPDS_04.1]
    And escalation timer T_ACK_3 starts with duration <T_ACK_3_MIN> minutes [Req_CPDS_04.3]

    And if (<SOC_TRACTION_PCT> < <BAT_GUARD_PCT> or <SOC_12V_PCT> < <BAT_GUARD_PCT>)
         and (<SOC_TRACTION_PCT> >= <SOC_CRIT_PCT> and <SOC_12V_PCT> >= <SOC_CRIT_PCT>)
      Then HVAC shall prioritize minimal energy cooling or ventilation
           while keeping CPDS in S5 and continuing the escalation timeline [Req_CPDS_04.2]
  # Additional scenarios for acknowledgment, escalation to carers, and critical battery
  # override are omitted here for brevity and are available in the online artifact.
\end{lstlisting}
\end{figure*}

\subsection{Test Script Generation}

In the test script generation step, we used the GPT-4.1 LLM to transform the previously generated Gherkin scenario with VSS paths into executable Python test artifacts.
The input to this step consists of three elements: (i) a Gherkin scenario as a structured behavioral specification that captures the intended test logic (Listing~\ref{lst:gherkin_scenario}), (ii) an example of the KUKSA client API in Python, and (iii) a minimal Behave framework example.

The KUKSA client API example demonstrated the synchronous simplified API from the \texttt{kuksa-python-sdk} and served as a template for generating code to retrieve and modify VSS values.
We deliberately chose the KUKSA client instead of the Velocitas framework (commonly used in the digital.auto playground) because direct modification of sensor values was essential for simulating the test conditions.
Furthermore, the synchronous API was sufficient for our purposes, as it avoided the additional complexity associated with asynchronous interactions.

The Behave framework example showed how to structure the test environment and step definitions.
This input illustrated how test steps can be organized into \texttt{environment.py} and \texttt{steps/*.py} files and how the VSS client fixture can be integrated with Behave lifecycle hooks.

The three inputs were combined into a structured prompt template for the LLM. The system message instructed the model to act as an automotive software test engineer, and the human message supplied the Gherkin scenario, the KUKSA client API example, and the Behave framework example. This combination allowed the LLM to map high-level behavioral specifications into code while adhering to established usage patterns of both the KUKSA API and the Behave framework. A simplified representation of the prompt is shown in Listing~\ref{lst:prompt_template}.
\begin{lstlisting}[style=prompt-style,
  caption={Prompt used for test code generation.}, label={lst:prompt_template}]
System message:
You are an automotive software test engineer. Your task is to write a Python script that will follow the provided Gherkin scenario and implement the required functionality using the synchronous KUKSA client API and the Behave framework. The script should be split into environment and steps files, including necessary imports and method definitions.
Human message:
--- Gherkin test case definition ---
{gherkin_test_case}
--- KUKSA Client API Example ---
{kuksa_client_api_example}
--- Behave Framework Example ---
{behave_example}
\end{lstlisting}
The output produced by the LLM was manually reviewed for correctness and completeness before integration into the test project. The Gherkin scenario was copied into the corresponding \texttt{.feature} file, the generated environment setup code was placed in \texttt{environment.py}, and the step definitions implementing the Given/When/Then clauses were written into \texttt{steps/feature\_steps.py}. The complete environment setup and step definition files are provided in digital.auto. Listing~\ref{lst:generated_step_definitions} shows a representative excerpt of the generated step definitions.

In earlier experiments with GPT-4o on the same prompt template, the generated code required manual addition of requirement identifiers in the decorator strings (for example, \texttt{[Req\_CPDS\_04.1]}), and some step names had to be aligned with the feature file. This limitation was not observed in the experiments reported in this paper, which use GPT-5 for the final test script generation.

\begin{figure*}[t]
\centering
\begin{lstlisting}[style=python-style,
  basicstyle=\ttfamily\scriptsize, 
  caption={Generated step definitions code (excerpt).}, label={lst:generated_step_definitions}]
# Given steps
@given('Vehicle.Cabin.ChildPresenceDetection.IsChildDetected is true [Req_CPDS_01.6]')
def step_impl(context):
    context.client.set_current_values({
        'Vehicle.Cabin.ChildPresenceDetection.IsChildDetected': Datapoint(True), })
@given('Vehicle.Cabin.Infotainment.DriverAppNotification.IsDriverNotified is true [Req_CPDS_04.1]')
def step_impl(context):
    context.client.set_current_values({
        'Vehicle.Cabin.Infotainment.DriverAppNotification.IsDriverNotified': Datapoint(True), })
\end{lstlisting}
\end{figure*}

\subsection{Test Execution}
For test execution, we first used digital.auto as a SiL platform and then ran the same generated artifacts in a real vehicle as a Vehicle-in-the-Loop (ViL) setup.
In the SiL configuration, a local instance of the SDV-Runtime and the KUKSA data broker run inside Docker and are connected to a multi-file project in the digital.auto playground. The project contains both the manually implemented prototype feature and the Behave-based test suite generated by Req2Road. The prototype implementation intentionally covers only the subset of CPDS logic required to exercise the HVAC adjustment scenario, allowing us to focus on end-to-end test generation and execution. An earlier version of this setup required several platform-specific workarounds (approximated VSS signals, single-file execution, and a custom BDD interpreter)~\cite{tum_sym_2025}. 
These constraints have since been removed in collaboration with the digital.auto engineering team, and the current experiments use a standard Behave runner without additional glue code. The SiL execution log in Listing~\ref{lst:prototype_execution_output} confirms that the prototype responds to changes in the relevant VSS signals and returns to standby after the HVAC intervention, as specified by the Gherkin scenario. 

\begin{figure*}[b]
\centering
\begin{lstlisting}[style=console-style,
  caption={Excerpt of digital.auto output after SiL test execution.}, label={lst:prototype_execution_output}]
APPL:     HVAC intervention was active for less than 5 minutes, moving to standby
APPL:     Child presence detection moved to standby
Feature: CPDS HVAC intervention # hvac_adjustment.feature:1
  Scenario: HVAC adjustment intervention (Req_CPDS_04)  # hvac_adjustment.feature:3
    Given Vehicle.Cabin.ChildPresenceDetection.IsChildDetected is true [Req_CPDS_01.6]
    ...
    Then Vehicle.Cabin.ChildPresenceDetection.IsEscalationActive is reset to false [Req_CPDS_04.2]
1 feature passed, 0 failed, 0 skipped
1 scenario passed, 0 failed, 0 skipped
8 steps passed, 0 failed, 0 skipped, 0 undefined
\end{lstlisting}
\end{figure*}

For ViL testing, we reuse the same Gherkin feature and generated Python step definitions in a production vehicle. The Behave test runner executes on a laptop that connects to an NVIDIA Jetson installed in the trunk; the Jetson aggregates in-vehicle communication and exposes the relevant signals as VSS paths through a KUKSA instance. Apart from updating the connection parameters from the digital.auto endpoint to the Jetson-based broker, no changes to the generated test logic are required. This shows that the artifacts can be transferred from SiL to ViL with minimal environment-specific adaptation.

\begin{figure}[t]
    \centering
    \includegraphics[width=.8\linewidth]{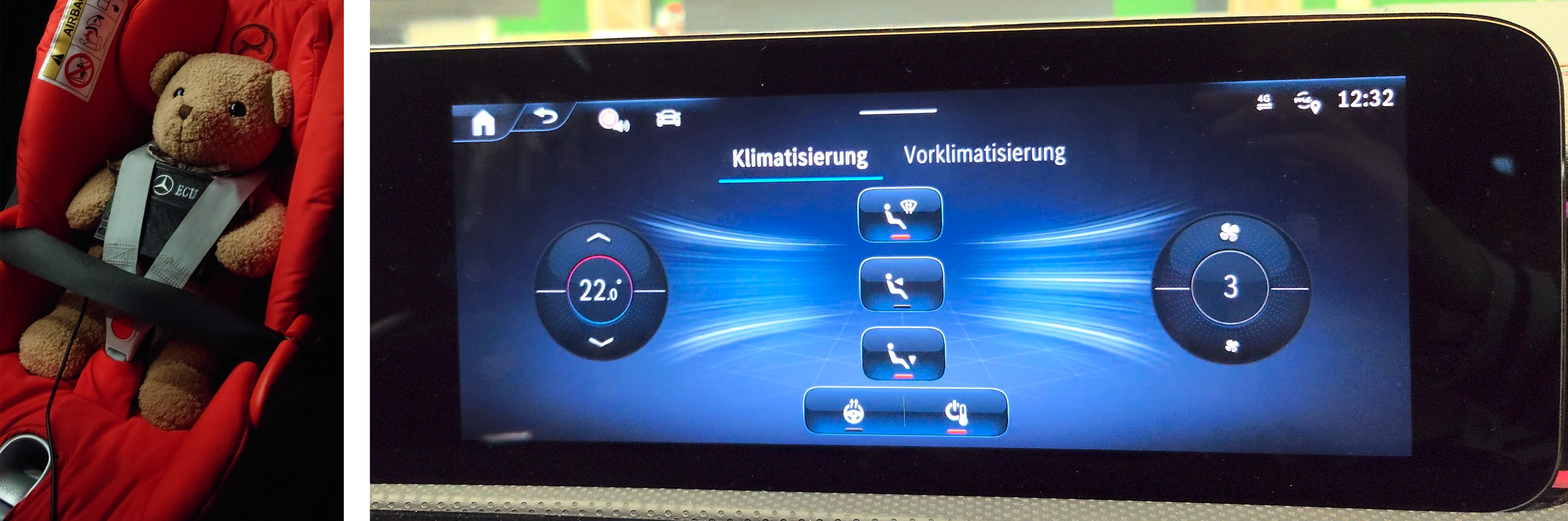}
    \caption{ViL setup: child seat with instrumented surrogate occupant (left) and corresponding HVAC state in the infotainment display (right).}
    \label{fig:vil_setup}
\end{figure}
To emulate child presence in an ethically acceptable but realistic way, we install a child restraint system with an instrumented surrogate occupant on the rear seat (Figure~\ref{fig:vil_setup}). A compact ECU attached to the harness plays prerecorded infant-crying audio and generates vibration patterns, while the surrounding sensors and in-vehicle signals are captured via the Jetson setup. During ViL execution, the generated test scripts trigger the HVAC adjustment intervention and then transition the CPDS to standby. The infotainment display confirms that the automatic HVAC override increases the cabin temperature from 18\,\textcelsius{} to 22\,\textcelsius{}, and an accompanying video (provided as supplementary material) shows the end-of-escalation sequence, including exterior light activation, door unlocking, and HVAC changes. The ViL experiments demonstrate that GenAI-generated test artifacts can drive and verify CPDS behavior on an actual vehicle. 

\section{Conclusion}
\label{discussion}
We introduced Req2Road, a GenAI-based pipeline that turns requirements and behavioral models into executable SDV tests.
Representative LLMs and VLMs were evaluated on three key tasks (Gherkin scenario generation, VSS signal mapping, and Python test script generation), instantiated the pipeline for a realistic CPDS use case, and executed the resulting tests both in the digital.auto playground and in a real vehicle.
The running example focused on Intervention I (HVAC adjustment) within a set of 36 natural-language constraints, using Gherkin and VSS paths as a unifying specification layer.
The primary contribution is an architectural demonstration of how LLMs, VLMs, and retrieval can be governed and composed to translate heterogeneous requirements artifacts into runnable tests across SiL and ViL.

Overall, the experiments show that, in a human-in-the-loop setting, GenAI can already produce test artifacts precise enough to exercise safety-relevant behavior in an SDV context. For test case generation, our GPT-4 class model with Chain-of-Thought and Few-Shot prompting produced executable Gherkin scenarios for 32 of 36 requirements (89\%) without modification; reviewers mainly corrected ambiguous edge cases and requirements with underspecified OR conditions. Given the feasibility scope, we report productivity impact as indicative and conditional on reviewer capabilities and requirement quality. A controlled comparison against manual and model-based approaches is future work.

VLMs, such as Gemini 2.5 Pro and Qwen 2.5 VL 72B, proved effective at grounding state- and transition-level signals in UML diagrams.
RAG-based VSS signal mapping with a prefiltered candidate list, using GPT-4o-mini, substantially improved selection quality compared to direct prompting over the full catalog. 
For code generation, the structured prompt, which combines Gherkin, KUKSA examples, and Behave templates, enabled the LLM to generate environment setup and step definitions that required only minimal manual adjustments.
These artifacts can be executed unchanged in SiL and ViL setups, apart from configuring the KUKSA endpoint.

The following discussion summarizes feasibility boundaries, remaining gaps, and how Req2Road integrates with existing test-generation approaches.
The case study suggests that Req2Road is feasible when (i) requirements can be grounded in observable vehicle signals, (ii) the signal vocabulary is stable enough to support mapping (here: VSS), and (iii) behavioral diagrams are legible and sufficiently structured for VLM extraction. Conversely, feasibility degrades for ambiguous requirements (especially OR-compositions), for large unfiltered signal catalogs, and for multi-requirement scenarios where code generation and oracle logic become brittle; these boundary conditions are reflected in the observed mapping degradation for 982 signals and the pass@k drop for three combined requirements. Remaining gaps therefore concern scalability (larger requirement sets and catalogs), stronger generation controls (schema validation and catalog validation), and safety assurance when generated tests influence release decisions.
Req2Road can complement existing test-generation approaches by acting as a translation layer between heterogeneous requirements artifacts and executable assets: model-based testing can provide structured behavioral models and coverage targets, while Req2Road can generate and maintain executable specifications (Gherkin) with explicit signal bindings (VSS) and test harness code for specific benches. For safety-critical contexts, Req2Road should be used with approval gates and guardrails aligned to pipeline stages (prompt and model-version logging, schema validation for generated Gherkin, catalog validation for VSS paths, execution sandboxing for generated code, and mandatory human approval for release-relevant tests); tool qualification and governance remain future work.

The evaluation highlights several limitations and failure modes that temper these positive results.
First, all LLMs struggled when confronted with large, unfiltered VSS catalogs: mapping quality degraded sharply when moving from 16 curated candidates to the full set of 982 signals, and smaller local models introduced many false positives.
This confirms that retrieval and prefiltering are not optional but necessary controls to curb hallucinations and keep the mapping task tractable.
Second, the code generation experiments show that correctness is sensitive to scenario complexity.
While all evaluated models achieved pass@1 equal to 1 for a single requirement, GPT-4o-mini and GPT-5 failed to produce any fully correct solutions when three requirements were combined, and only GPT-4.1 reached pass@3 equal to 1 in that setting.
In practice, this means that for richer scenarios, we cannot simply prompt once and trust the first answer; sampling, execution-based filtering, and human review remain essential.

Third, several constraints stem from the experimental setup and scope.
In our prior work~\cite{tum_sym_2025}, we considered a smaller and simpler HVAC-only use case with a limited set of requirements.
In contrast, Req2Road covers the full CPDS escalation logic, including states and edge cases.
This broader scope made correctness checking noticeably more demanding and often required clarifications of vague or intertwined requirements, reinforcing the need for systematic structuring and clarification of natural-language requirements before automated test generation.
Similarly, the SiL and ViL evaluations still target a single vehicle configuration and a single SDV platform.
Although the ViL experiments confirm that the generated artifacts can be transferred from digital.auto to an on-vehicle Jetson-based setup with minimal changes, additional studies across different vehicle architectures, ECUs, and signal catalogs are needed to assess robustness.
Finally, data protection and governance concerns limit the use of commercial models: while they are suitable for downstream artifacts such as refined Gherkin and code, direct prompting on proprietary requirements should rely on locally deployable LLMs and VLMs, which currently lag slightly behind in accuracy.

Fourth, our evaluation primarily reports artifact executability and mapping correctness rather than fault detection capability. Tests can be executable yet ineffective if assertions are weak or negative cases are missing. Oracle correctness is a key risk because expected values and timing constraints may be incomplete or incorrectly inferred from requirements. Future work will assess effectiveness via fault seeding or mutation testing (e.g., threshold inversions, timing shifts, missing escalation steps) and strengthen oracles using metamorphic relations and differential checks across SiL and ViL.

Despite these limitations, the study indicates that a human-in-the-loop GenAI pipeline can feasibly translate heterogeneous requirements artifacts into runnable tests across SiL and ViL settings; quantifying productivity effects across roles and projects remains future work. Req2Road illustrates how LLMs and VLMs can be combined with RAG, VSS catalogs, and SDV runtimes to reduce the gap between textual specifications, signal-level models, and executable tests. At the same time, the observed failure modes point to further research needs, including better support for logical compositions in requirements, automated consistency checks across generated artifacts, stronger guarantees for code generation in safety-critical contexts, and systematic validation on larger requirement sets and diverse SDV platforms. 

\noindent\textbf{{\ackname}} This work has received funding from the European Chips JU under Framework Partnership Agreement No 101139789 (HAL4SDV), including the national funding from the Federal Ministry of Research, Technology, and Space of Germany under grant number 16MEE00471K.
\bibliographystyle{splncs04}
\bibliography{ref.bib}

\end{document}